\documentclass{article}
\usepackage{spconf,amsmath,graphicx}
\usepackage{cite}
\usepackage{amssymb,amsfonts}
\usepackage{algorithmic}
\usepackage{textcomp}
\usepackage{subcaption}
\usepackage[table,xcdraw]{xcolor}
\usepackage{tcolorbox}
\usepackage{soul} 
\usepackage{etoolbox}
\usepackage{array, makecell} 
\apptocmd{\thebibliography}{\small}{}{}
\usepackage{hyperref}
\newcommand{\sdrlap}[1]{%
  \rlap{%
    \kern0.15em%
    \textcolor{blue}{%
      \fontsize{2pt}{2pt}\selectfont(#1)%
    }%
  }%
}

\newcommand{\pct}[1]{%
  \rlap{%
    \kern0.12em
    \textcolor{blue}{\fontsize{4pt}{4pt}\selectfont(#1)}%
  }%
}

\setul{0.18ex}{0.15ex}

\title{TRIQA: Image Quality Assessment By Contrastive Pretraining on Ordered Distortion Triplets}
\name{Rajesh Sureddi$^{*}$, Saman Zadtootaghaj$^{\dagger}$, Nabajeet Barman$^{\dagger}$, Alan C. Bovik$^{*}$}
\address{The University of Texas at Austin$^{*}$, Sony Interactive Entertainment$^{\dagger}$}
\begin{document}
\maketitle
\begin{abstract}
Image Quality Assessment (IQA) models aim to predict perceptual image quality in alignment with human judgments. No-Reference (NR) IQA remains particularly challenging due to the absence of a reference image. While deep learning has significantly advanced this field, a major hurdle in developing NR-IQA models is the limited availability of subjectively labeled data. Most existing deep learning-based NR-IQA approaches rely on pre-training on large-scale datasets before fine-tuning for IQA tasks. To further advance progress in this area, we propose a novel approach that constructs a custom dataset using a limited number of reference content images and introduces a no-reference IQA model that incorporates both content and quality features for perceptual quality prediction. Specifically, we train a quality-aware model using contrastive triplet-based learning, enabling efficient training with fewer samples while achieving strong generalization performance across publicly available datasets. Our repository is available at \url{https://github.com/rajeshsureddi/triqa}.
\footnote{We thank the Texas Advance Computing Center and the National Science Foundation AI Institute for Foundations of Machine Learning (Grant 2019844) for providing compute resources that contributed to our research.}

\end{abstract}
\section{Introduction}
\label{sec:intro}

With the increasing use of smartphones, digital cameras, and other electronic devices, the daily production, streaming, and sharing of images has soared. Billions of images are shared over the internet daily, particularly on social platforms like Meta, Twitter (X), and LinkedIn. It is crucial to ensure that these images are posted without compromising their quality, as visual appeal is important to viewers. The process of evaluating how closely an image aligns with human visual perception is known as image quality assessment (IQA). IQA methods are categorized based on the availability of a reference image. Full-reference IQA (FR-IQA)involves both a presumably pristine reference image and a corresponding distorted image, while reduced-reference IQA uses only partial information from a reference image. No-reference IQA (NR-IQA), on the other hand, operates without access to any reference image. Since streaming companies and social media platforms deliver content often lacking a reference image, there is a significant demand in the research community for improved NR-IQA models. Early successful NR-IQA models \cite{mittal2012making},\cite{saad2012blind} were developed by measuring stastical perturbations of distorted images from bandpass natural scene models, often trained on datasets of distorted images and mean opinion scores (MOS). More recently, deep learning-based IQA models \cite{zhang2018blind,su2020blindly,golestaneh2022no,ke2021musiq,ying2020patches,madhusudana2022image,saha2023re} have been developed that compute and inference using a variety of evolving architectures and loss functions. Most of these approaches aim to extract either content and quality features together, or separately and then combine them. Likewise, our model extracts quality-and content-aware features separately, then combines them afterwards. Specifically, we deploy a pre-trained ConvNeXt\cite{liu2022convnet} backbone trained on ImageNet, that extracts content aware features and we introduce a novel training triplet-based strategy for extracting quality-aware features. While some quality-aware models \cite{madhusudana2022image,saha2023re,agnolucci2024arniqa} have used contrastive learning to learn feature representations, they have not incorporated relative rankings of distorted training images. We demonstrate the efficacy of contrastive learning on: triplets of training images processed by diverse levels of each of several synthetic distortions.
Thus, both the content-aware and quality-aware backbones are pretrained without MOS, hence are unsupervised against human subjecting. While our approach designed for NR-IQA, it can be adapted to FR-IQA tasks without the need for additional training or fine-tuning.

\begin{figure*}[h!]

    \begin{minipage}{\textwidth}
        \begin{minipage}[b]{0.161\textwidth}
            \includegraphics[width=\textwidth]{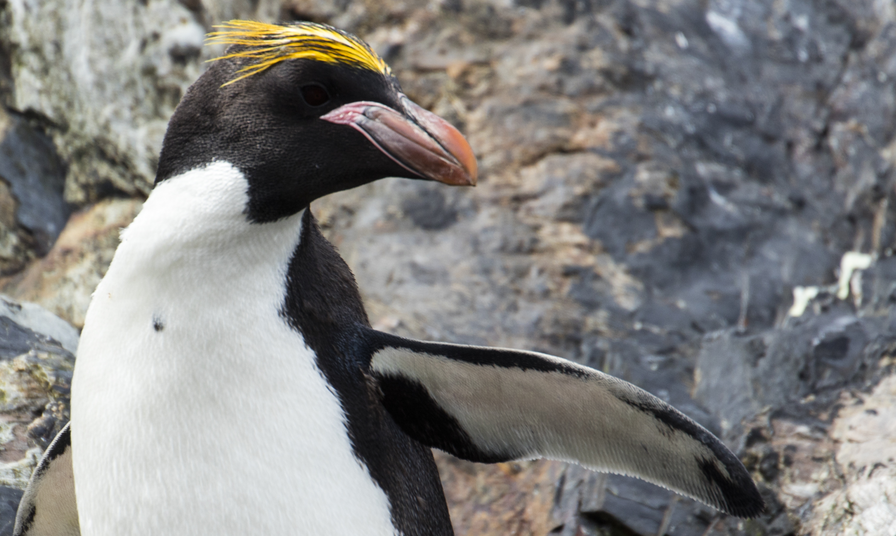}
            \caption*{\scriptsize Pristine}
        \end{minipage}
        \begin{minipage}[b]{0.161\textwidth}
            \includegraphics[width=\textwidth]{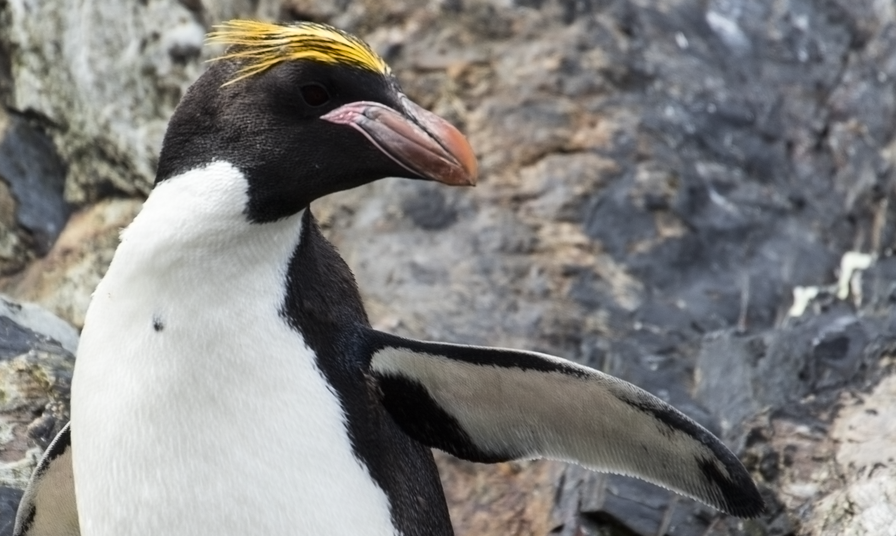}
            \caption*{\scriptsize d0}
        \end{minipage}
        \begin{minipage}[b]{0.161\textwidth}
            \includegraphics[width=\textwidth]{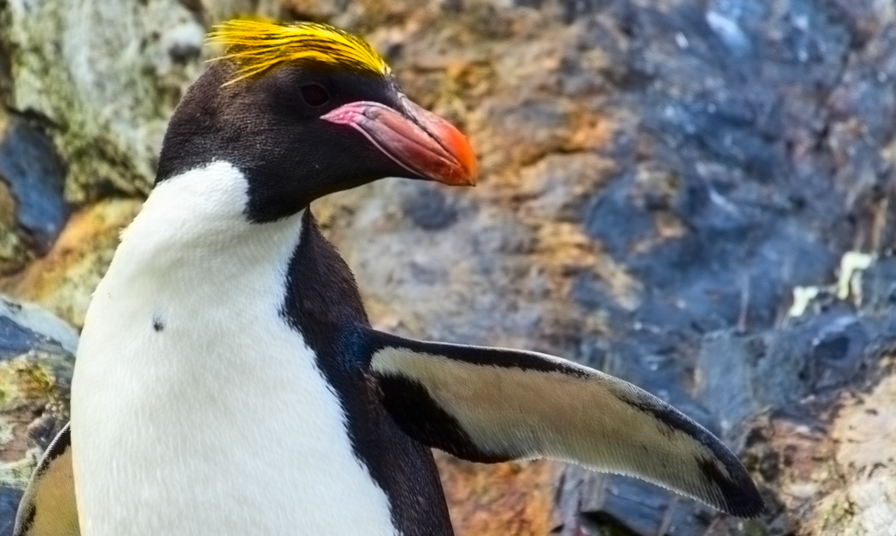}
            \caption*{\scriptsize d1}
        \end{minipage}
        \begin{minipage}[b]{0.161\textwidth}
            \includegraphics[width=\textwidth]{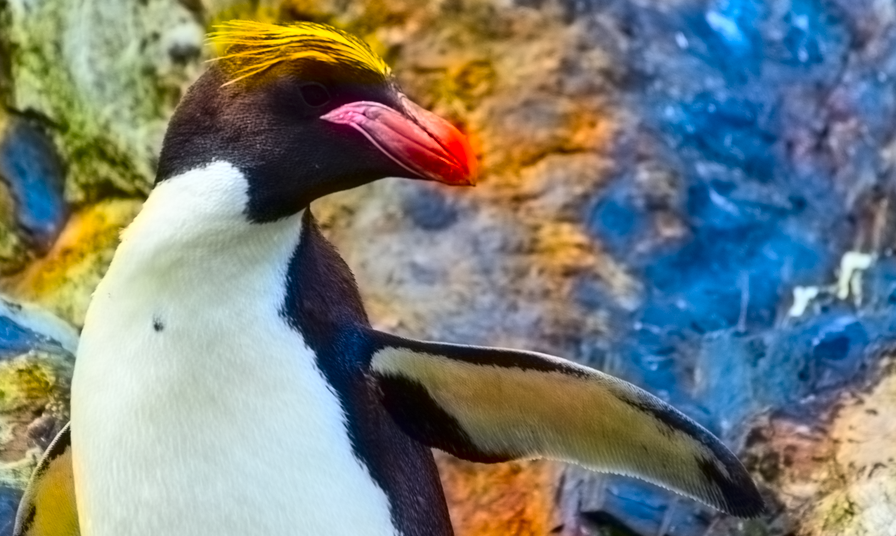}
            \caption*{\scriptsize d2}
        \end{minipage}
        \begin{minipage}[b]{0.161\textwidth}
            \includegraphics[width=\textwidth]{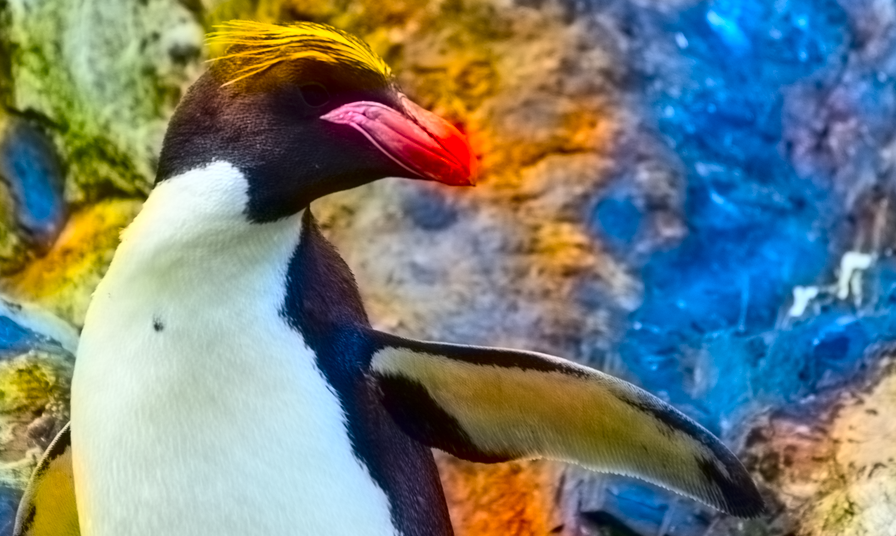}
            \caption*{\scriptsize d3}
        \end{minipage}
        \begin{minipage}[b]{0.161\textwidth}
            \includegraphics[width=\textwidth]{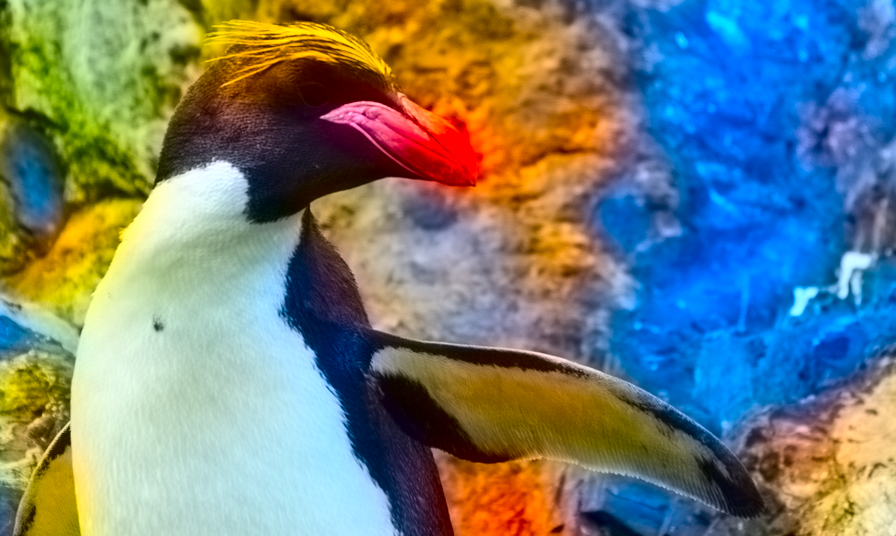}
            \caption*{\scriptsize d4}
        \end{minipage}
    \end{minipage}
    
    \vspace{0.1cm} 
    \begin{tcolorbox}[colback=gray!10, colframe=gray!50, boxrule=0.1mm, width=\dimexpr\textwidth\relax, sharp corners, right=-15.5cm, height=8cm]
    \begin{minipage}{\textwidth}
        \begin{minipage}[b]{0.08\textwidth}
            \includegraphics[width=\textwidth]{images/crop256x256/0801_imcolordiffuse_0.png}
            \caption*{\scriptsize d0}
        \end{minipage}
        \begin{minipage}[b]{0.08\textwidth}
            \includegraphics[width=\textwidth]{images/crop256x256/0801_imcolordiffuse_1.png}
            \caption*{\scriptsize d1}
        \end{minipage}
        \begin{minipage}[b]{0.08\textwidth}
            \includegraphics[width=\textwidth]{images/crop256x256/0801_imcolordiffuse_2.png}
            \caption*{\scriptsize d2}
        \end{minipage}
        \hspace{0.5cm}
        \begin{minipage}[b]{0.08\textwidth}
            \includegraphics[width=\textwidth]{images/crop256x256/0801_imcolordiffuse_1.png}
            \caption*{\scriptsize d1}
        \end{minipage}
        \begin{minipage}[b]{0.08\textwidth}
            \includegraphics[width=\textwidth]{images/crop256x256/0801_imcolordiffuse_2.png}
            \caption*{\scriptsize d2}
        \end{minipage}
        \begin{minipage}[b]{0.08\textwidth}
            \includegraphics[width=\textwidth]{images/crop256x256/0801_imcolordiffuse_3.png}
            \caption*{\scriptsize d3}
        \end{minipage}
        
        \vspace{0.2cm} 
        
        \begin{minipage}[b]{0.08\textwidth}
            \includegraphics[width=\textwidth]{images/crop256x256/0801_imcolordiffuse_0.png}
            \caption*{\scriptsize d0}
        \end{minipage}
        \begin{minipage}[b]{0.08\textwidth}
            \includegraphics[width=\textwidth]{images/crop256x256/0801_imcolordiffuse_1.png}
            \caption*{\scriptsize d1}
        \end{minipage}
        \begin{minipage}[b]{0.08\textwidth}
           \includegraphics[width=\textwidth]{images/crop256x256/0801_imcolordiffuse_3.png}
            \caption*{\scriptsize d3}
        \end{minipage}
        \hspace{0.5cm}
        \begin{minipage}[b]{0.08\textwidth}
            \includegraphics[width=\textwidth]{images/crop256x256/0801_imcolordiffuse_1.png}
            \caption*{\scriptsize d1}
        \end{minipage}
        \begin{minipage}[b]{0.08\textwidth}
            \includegraphics[width=\textwidth]{images/crop256x256/0801_imcolordiffuse_2.png}
            \caption*{\scriptsize d2}
        \end{minipage}
        \begin{minipage}[b]{0.08\textwidth}
            \includegraphics[width=\textwidth]{images/crop256x256/0801_imcolordiffuse_4.png}
            \caption*{\scriptsize d4}
        \end{minipage}
        
        \vspace{0.2cm} 
        
        \begin{minipage}[b]{0.08\textwidth}
            \includegraphics[width=\textwidth]{images/crop256x256/0801_imcolordiffuse_0.png}
            \caption*{\scriptsize d0}
        \end{minipage}
        \begin{minipage}[b]{0.08\textwidth}
            \includegraphics[width=\textwidth]{images/crop256x256/0801_imcolordiffuse_1.png}
            \caption*{\scriptsize d1}
        \end{minipage}
        \begin{minipage}[b]{0.08\textwidth}
            \includegraphics[width=\textwidth]{images/crop256x256/0801_imcolordiffuse_4.png}
            \caption*{\scriptsize d4}
        \end{minipage}
        \hspace{0.5cm}
        \begin{minipage}[b]{0.08\textwidth}
            \includegraphics[width=\textwidth]{images/crop256x256/0801_imcolordiffuse_2.png}
            \caption*{\scriptsize d2}
        \end{minipage}
        \begin{minipage}[b]{0.08\textwidth}
            \includegraphics[width=\textwidth]{images/crop256x256/0801_imcolordiffuse_3.png}
            \caption*{\scriptsize d3}
        \end{minipage}
        \begin{minipage}[b]{0.08\textwidth}
            \includegraphics[width=\textwidth]{images/crop256x256/0801_imcolordiffuse_4.png}
            \caption*{\scriptsize d4}
        \end{minipage}
        
    \end{minipage}

\end{tcolorbox}
\caption{Examples of triplet formulation using samples from DIV2K\cite{Agustsson_2017_CVPR_Workshops} dataset. The top row displays a pristine image alongside five progressively distorted versions of it (d0-d4). The rows beneath present the rank-ordered triplets created from these images.}
\label{triplets}
\end{figure*}

\section{Related work}

Early successfull NR-IQA models like NIQE\cite{mittal2012making}, BLIINDS\cite{saad2012blind}, and PIQUE\cite{venkatanath2015blind} extracted hand-crafted features expressive of deviations of image naturalness. Recent deep learning approaches, however, have evolved to automatically extract features that are sensitive to image quality. For example, RankIQA\cite{Liu_2017_ICCV} is a siamese network trained to rank pairs of images,  then fine-tuned on the synthetic LIVE\cite{sheikh2006statistical} and TID\cite{mikhailiuk2018cross} datasets. However, this approach was limited to generating ranked pairs altered by only a few synthetic distortions, and required separate pre-training on each fine-tuning task, reducing generalization. DB-CNN \cite{zhang2018blind} uses two CNNs: one for synthetic distortions and another for authentic distortions. The CNN for synthetic distortions was pre-trained to classify distortion types and levels, while the CNN for authentic distortions used features from a pre-trained image classification network. The combined model was then fine-tuned on databases using the $l2$ loss. HyperIQA\cite{su2020blindly} is a supervised learning approach that first learns semantic features extracted by a pre-trained network, then feeds them to a  Hyper Network that predicts image quality by weighting and aggregating multi-scale features from the content network, representative of both local and global distortions. PaQ-2-PiQ\cite{ying2020patches} extracts both local and global quality-aware features using a large patch-labeled images quality database. TReS \cite{golestaneh2022no} employs CNNs and Transformers to model local and non-local features, training on relative ranking to capture image correlations. However, this method still requires training a separate model for each dataset, limiting generalizability. MUSIQ\cite{ke2021musiq} analyzes each entire image at multiple scales to extract features, then pools the obtained representations using Vision Transformers to predict quality scores. CONTRIQUE\cite{madhusudana2022image} deploys a self-supervised approach where by a deep CNN model is trained using contrastive learning, posing distortion type and degree as auxiliary tasks to learn representations that are then used to train a regression model that predicts quality scores. ReIQA\cite{saha2023re} improves upon CONTRIQUE\cite{madhusudana2022image} using a mixture of expert approach, combining quality and content-aware features that are trained independently, along with an engineered augmentation method for better representation learning. ARNIQA\cite{agnolucci2024arniqa} deploys an image degradation model to train a SimCLR\cite{chen2020simple} network, maximizing similarities between tiles from different images to learn distortion-related manifolds, unlike Re-IQA, which utilizes on within-image crops. 

All the deep-learning methods just described are able to produce very accurate and generalizable NR-IQA predictions provided they have access to large pretraining and/or training datasets. Our approach differs by using a limited number of reference images while leveraging a large set of diverse image impairments, ranked in triplets by severity, to train our quality-aware branch. This minimizes content dependency, which is already handled by the content-aware branch, enabling more efficient learning with fewer reference images while focusing on extracting quality-related features.

\begin{figure}[ht]
    \centering
    \begin{minipage}[b]{0.12\textwidth}
        \centering
        \includegraphics[width=\textwidth]{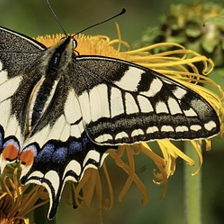}
        \caption*{(a)}
    \end{minipage}
    \hspace{6mm}
    \begin{minipage}[b]{0.12\textwidth}
        \centering
        \includegraphics[width=\textwidth]{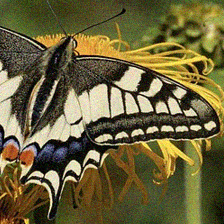}
        \caption*{(b)}
    \end{minipage}
    \hspace{6mm}
    \begin{minipage}[b]{0.12\textwidth}
        \centering
        \includegraphics[width=\textwidth]{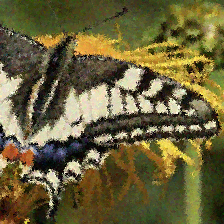}
        \caption*{(c)}
    \end{minipage}
    \caption{An example of a triplet illustrating combinations of distortions. (a) Pristine image; (b) Image after applying \textit{Gaussian noise}; (c) Image after applying \textit{Gaussian noise} and \textit{Jitter}. The differences are more clearly visible in the zoomed-in display.}
    \label{combination_triplet}
\end{figure}

\section{Method}
\label{Method}

\begin{figure*}[ht!]
    \centering
    \includegraphics[width=0.485\textwidth,height=0.18\textheight]{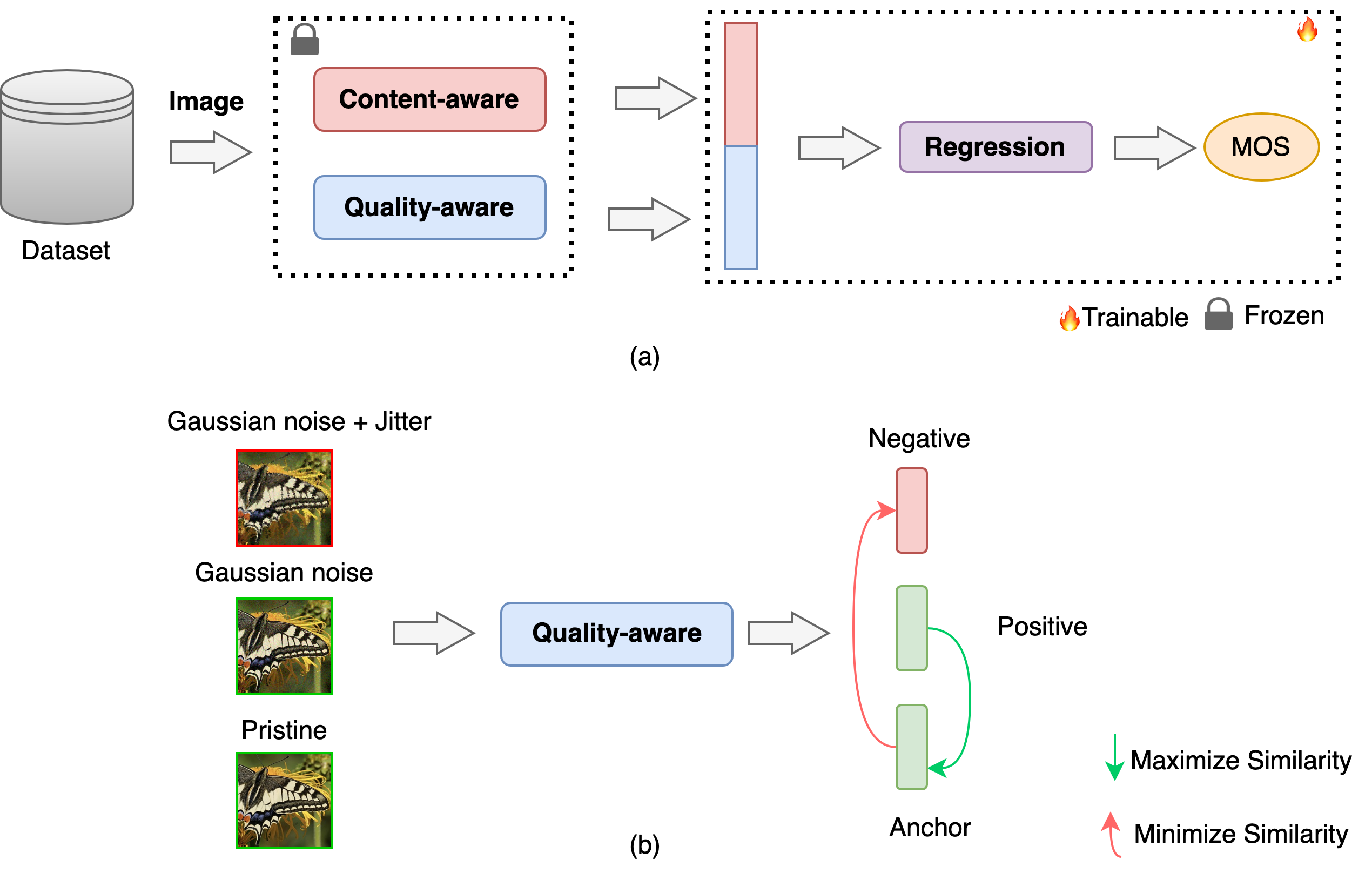}
    \caption{(a) Flow diagram of the TRIQA quality assessment model, (b) Illustration of the training methodology for the quality-aware network. For improved visual quality, please zoom-in.}
    \label{pipeline}
\end{figure*}

As mentioned, our model, which we call TRIQA, combines content-and quality-aware representations using a simple linear regression network to predict the quality score. As shown in Fig \ref{pipeline} (a), we deploy a ConvNeXt\cite{liu2022convnet} backbone originally trained for image classification on the ImageNet-1K dataset. ConvNeXt is a modified ResNet architecture that is able to compete with Vision Transformers (ViT) architectures on image classification \cite{liu2022convnet} and self-supervised \cite{woo2023convnext} tasks. The second backbone uses the same pretrained ConvNeXt, but to learn quality representations using fine tuned a novel training strategy that requires many fewer training content image samples than other models.

\subsection{\textit{Training Data}}
\label{Training Data}
We used the DIV2K \cite{Agustsson_2017_CVPR_Workshops} dataset to create the triplets using 20 synthetic distortions, with five levels of degradations as provided in KADID-10k\cite{kadid10k} dataset. We generated triplets by ranking them by increasing distortion severity, as shown in Fig.~\ref{triplets}. We denote each triplet as $[anchor, positive, negative]$ where \textit{anchor} is the least degraded or reference image, \textit{negative} is most degraded, and \textit{positive} is degraded more than the \textit{anchor} image and less than the \textit{negative} image. Assuming a pristine image and five levels of distortion as depicted in Fig.~\ref{triplets}, generate all 20 unique ordered triplets. Fig.~\ref{triplets} only depicts the six out of 10 triplets created from the distorted images only. Another 10 triplets are obtained using the pristine image as the first triplet element. To better approach the very general nature of real-world distorted images we also created another set of triplets of images impaired by multiple, combined synthetic distortions. The approach taken in ARNIQA grouped similar distortions, then created combinations of multiple distortions only within each group, excluding distortions from other groups. This strategy limits the number of possible combinations of distortions. Here we instead generate pairs of distortions that originate from different groups rather than from the same group. To accomplish this, we first identify all possible combinations of distortion groups. For each combination of groups, we examine all distortions within the first group and systematically pair them with every distortion in the second group. We then distortion the pristine images by applying the two distortions sequentially. This is done for varying levels of each distortion. This methodology facilitates the generation of large number of distortion pairs from a small number of pristine samples. For example, consider a sampled pair \([Gaussian\ noise, Jitter]\) drawn from different groups. From this pair, we can create the following triplets, as illustrated in Fig.~\ref{combination_triplet}:
\begin{align*}
&[\text{pristine, Gaussian noise\_d1, Gaussian noise\_d1\_Jitter\_d1}], \\
&[\text{pristine, Gaussian noise\_d1, Gaussian noise\_d1\_Jitter\_d3}], \\
&[\text{pristine, Gaussian noise\_d3, Gaussian noise\_d3\_Jitter\_d1}], \\
&[\text{pristine, Gaussian noise\_d3, Gaussian noise\_d3\_Jitter\_d3}].
\end{align*}

These triplets satisfy the relative ranking requirements, because applying either distortion level d1 or d3 first, followed by any level (d1 or d3), results in a more distorted image than applying a single level alone. Thus, we construct triplets by designating pristine images as anchors, using for example, the image distorted with d1 or d3 as the positive sample, then further distorting the positive sample with d1 or d3 as the negative sample.

To clarify, the number of triplets formed using only synthetic distortions across five levels is \({N \choose k}\), where \(N=6\) (including the reference image and the five distortion levels) and k=3 (triplets), resulting in 20 triplets for each distortion. If 20 different types of distortion are applied, then 400 triplets per image are obtained. Additionally, when combinations of multiple distortion pairs are included, 608 additional triplet combinations are obtained per image.
By applying this procedure to the 800 training and 100 validation images in DIV2K, we obtain a total of 806,400 training samples (comprising 320,000 general triplets and 486,400 triplet combinations) and 100,800 validation samples (including 40,000 general triplets and 60,800 triplet combinations).
The variety of distortions in the triplets enhances the model
ability to learn very general, quality-related features, on a very limited number of reference
contents.

\subsection{\textit{Evaluation Dataset}}
\label{Evaluation_Dataset}
We utilized several publicly available User-Generated Content (UGC) image quality datasets, including CLIVE \cite{ghadiyaram2015massive}, KonIQ \cite{koniq10k}, SPAQ \cite{fang2020cvpr}, and FLIVE. These datasets consist of images collected from user-generated image platforms, with subjective quality studies conducted either online or in laboratory settings. CLIVE contains 1,162 mobile-captured images, KonIQ includes 10,073 images collected in the wild, SPAQ has 11,125 smartphone-captured images, and FLIVE, the largest available dataset to date, contains 39,811 images. We also used synthetic distortion datasets to evaluate learned representations for the FR-IQA task. The  datasets LIVE-IQA \cite{sheikh2006statistical}, TID2013, CSIQ-IQA \cite{larson2010most}, and KADID10K contain reference images along with corresponding distorted versions. Specifically, LIVE-IQA comprises 779 images generated from distortions applied to 30 reference images, TID2013 includes 3,000 images created by distorting 25 reference images, and CSIQ-IQA holds 866 images derived from 30 references. KADID10K contains 10,125 images generated by distorting 81 reference images.  

\subsection{\textit{Training}}
\label{Training}

We use a ConvNeXt architecture and its pretrained weights to initialize the quality-aware network, reshaping the output of the last layer into a 128-dimensional vector. As shown in Fig.~\ref{pipeline} (b), we pass each triplet through the network to obtain a 128-dimensional feature representation for each sample, and a triplet margin loss function to learn the weights. Denoting [$x$,$x^+$,$x^-$] as having elements that are anchor, positive, and negative image samples, and denoting the quality network by $F$, and letting

\begin{equation}
    a = F(x), \quad p = F(x^+), \quad n = F(x^-),
\end{equation}
\\
then the triplet margin loss is 

\begin{equation}
    L(a,p,n) = \max \left\{ d(a, p) - d(a, n) + \text{margin}, \, 0 \right\}
\end{equation}
\\
\text{where}
$d(x, y) = \|x - y\|_2$ and where we fixed margin=1.5.
As mentioned in Section \ref{Training Data}, We use 806,400 training samples and 100,800 validation samples to validate the learned network during training. The model rapidly converged in both training and validation within a single epoch, so we saved it after the first epoch to extract quality-aware representations. We used the Adam optimizer with eps=1e-8, betas (0.9, 0.999), learning rate = 5e-4 and momentum=0.9, and the CosineLR scheduler during optimization. We generated RandomCrop coordinates to crop a $256 \times 256$ image from one sample in the triplet, and then used the same coordinates to crop all samples in the triplet.

\begin{table*}    
    \caption{Comparison of different IQA models on authentic UGC distortions. Bold values represent the best performance, underlined values indicate the second-best, and values in parentheses denote standard deviations.}
    \centering
    \scriptsize
    \tabcolsep=0.3cm
    \renewcommand{\arraystretch}{1.15}
    \resizebox{0.88\linewidth}{!}{
    \begin{tabular}{|l|l|rr|rr|rr|rr|}
        \hline
        &  & \multicolumn{2}{c}{\textbf{FLIVE}} 
             & \multicolumn{2}{c}{\textbf{SPAQ}} 
             & \multicolumn{2}{c}{\textbf{KonIQ}} 
             & \multicolumn{2}{c|}{\textbf{CLIVE}} \\ 
        Method & Type 
             & SRCC & PLCC 
             & SRCC & PLCC 
             & SRCC & PLCC 
             & SRCC & PLCC \\ \hline

        BRISQUE   & Handcrafted 
             & 0.288 & 0.373 
             & 0.809 & 0.817 
             & 0.665 & 0.681 
             & 0.608 & 0.629  \\ 
        NIQE      &            
             & 0.211 & 0.288 
             & 0.700 & 0.709 
             & -    & -    
             & -    & -    \\ \hline
    
        DB-CNN    & Supervised 
             & 0.554 & 0.652 
             & 0.911 & 0.915 
             & 0.875 & 0.884 
             & \ul{0.851} & 0.869  \\ 
        HyperIQA  &           
             & 0.535 & 0.623 
             & \ul{0.916} & \ul{0.919} 
             & 0.906 & 0.917 
             & \textbf{0.859} & \textbf{0.882} \\ 
        TReS      &           
             & 0.554 & 0.625 
             & -    & -    
             & -    & -   
             & -    & -    \\ \hline
        
        CONTRIQUE & SSL+LR   
             & 0.580 & 0.641 
             & 0.914 & \ul{0.919} 
             & 0.894 & 0.906 
             & 0.845 & 0.857 \\ 
        Re-IQA    &           
             & \textbf{0.645} & \textbf{0.733} 
             & \textbf{0.918} & \textbf{0.925} 
             & \ul{0.914} & \ul{0.923} 
             & 0.840 & 0.854 \\ 
        ARNIQA    &           
             & \ul{0.595} & \ul{0.671} 
             & 0.905 & 0.910 
             & -    & -    
             & -    & -    \\ 
    \rowcolor[gray]{0.9}
    \textbf{TRIQA} & 
      & 0.567 & 0.653 
      & 0.911\sdrlap{0.0033} & 0.914\sdrlap{0.0037} 
      & \textbf{0.915}\sdrlap{0.0036} & \textbf{0.926}\sdrlap{0.0018} 
      & 0.837\sdrlap{0.015} & \ul{0.871}\sdrlap{0.011} \\
        \hline
    \end{tabular}}
    \label{IQA_authentic_table}
\end{table*}

\begin{table*}
    \caption{Performances of FR-IQA without any fine-tuning or training on the dataset.}
	\centering
    \scriptsize
	\tabcolsep=0.3cm
	\renewcommand{\arraystretch}{1.15}
	\resizebox{0.88\linewidth}{!}{
    \begin{tabular}{|l |cc |cc |cc |cc | cc|}
        \hline
        & \multicolumn{2}{c}{\textbf{CSIQ}} & \multicolumn{2}{c}{\textbf{KADID}} & \multicolumn{2}{c}{\textbf{TID2013}} & \multicolumn{2}{c|}{\textbf{LIVE-IQA}} & \multicolumn{2}{c|}{\textbf{Average}}\\ 
        NR-IQA Models implemented as FR-IQA models & SRCC & PLCC & SRCC & PLCC & SRCC & PLCC & SRCC & PLCC & SRCC & PLCC\\ \hline

        CONTRIQUE &0.676& 0.696&0.613& 0.618&0.450&0.460&
0.802& \ul{0.795} & 0.635& 0.642\\ 
        Re-IQA &0.691& 0.518&0.547&0.492&0.398&0.397&0.662& 0.554 &0.575 & 0.490\\ 
        ARNIQA&\ul{0.853}&\ul{0.844}&\ul{0.742}&\textbf{0.731}&\textbf{0.604}&\ul{0.637}&\ul{0.866}& \textbf{0.857} & \ul{0.766}& \textbf{0.767}\\ 
        \rowcolor[gray]{0.9}
        \textbf{TRIQA-FR} & \textbf{0.918}&\textbf{0.855}&\textbf{0.773}&\ul{0.667}&\ul{0.589}&\textbf{0.689}&\textbf{0.895}& 0.782 & \textbf{0.794} & \ul{0.755}
\\

        \hline
    \end{tabular}}
    \label{fr_iqa_table}
\end{table*}

\subsection{\textit{Evaluation Methods}}

When learning NR IQA models on UGC datasets, we extracted features from the quality-aware branch at two scales: full and half. The pretrianed content-aware ConvNeXt features are combined with the quality-aware features to train a Support Vector Regressor (SVR), tuned via GridSearch to select the best configuration to fit a LinearSVR. Specifically, we extracted 1,536 features each from each network branch, merging them into 3,072-dimensional feature vectors (per image) which were used to train the SVR to predict MOS. Model performance is measured on IQA datasets using Spearman's rank correlation coefficient (SRCC) and Pearson's linear correlation coefficient (PLCC). We randomly split each data set used for regression into 80\% and 20\% train test holdouts, and the median performances over 10 iterations are reported in the following Tables. On FLIVE, which is the largest dataset, we report the results over just one iteration.

\section{Results}
This section compares the performance of TRIQA against state-of-the-art (SoTA) models in terms of SRCC and PLCC on the representative IQA datasets in Tables \ref{IQA_authentic_table} and \ref{fr_iqa_table}. On all the authentic UGC IQA datasets in Table \ref{IQA_authentic_table}, TRIQA delivered quality prediction performances on par with all the compared SoTA models, as shown in Table \ref{IQA_authentic_table}. In addition to the NR-IQA task, we aimed to evaluate how well the latent features of TRIQA were trained. To achieve this, we created a full-reference model without training on any quality dataset, measuring the cosine similarity of the embedding/latent features of TRIQA after excluding the content model. Table \ref{fr_iqa_table} highlights the cross-database robustness of this approach, referred to as TRIQA-FR, which achieved the most consistent results across all datasets and the highest average performance. TRIQA-FR extracts quality-aware features from pristine images in each synthetic dataset and compares them with the corresponding degraded images using cosine similarity to predict quality scores, without requiring fine-tuning or additional training. Similarly, other SoTA NR-IQA models were adapted by extracting features from their respective pre-trained models and applying the same cosine similarity function to reference and distorted images. The results confirm that TRIQA-FR effectively assesses quality based solely on its learned features, providing perceptual scores independent of subjective feedback. This demonstrates the effectiveness of triplet-learned representations in evaluating perceptual image quality relative to their references. It may be noted that using a similar number of features as CONTRIQUE, Re-IQA, and ARNIQA, TRIQA is highly competitive with them on UGC datasets and more genralized, and TRIQA-FR outperforms them on synthetic datasets for the FR-IQA task. This is achieved through the use of contrastive self-supervised learning (SSL) on distorted image triplets to learn quality-related features, combined with content-aware features via a linear regression (LR) head which maps them to quality predictions. Notably, TRIQA achieves this impressive performance using a significantly smaller number of original content samples - only 800 - as compared to the millions of samples required by other SoTA methods, for extracting quality related features.
\subsection{Ablation Study}
In this section, we present an evaluation comparing the performance of the quality-aware branch when trained with and without the combination of multiple distortion triplets, which were created as described in Section \ref{Training Data}. Our analysis reveals that including distortion triplets during training significantly boosts performance compared to excluding them as shown in Table \ref{ablation_table}. This demonstrates the capability of the quality-aware model to effectively extract features related to user-generated distortions, highlighting its promise in handling such challenges.

\begin{table}
    \caption{Performance evaluation of the quality-aware model on authentic UGC distortions, comparing results with and without multiple distortion triplets included during training (percentages indicate improvements).}
    \centering
    \scriptsize
    \tabcolsep=0.05cm
    \resizebox{0.9\linewidth}{!}{
    \begin{tabular}{|l|cc|cc|}
        \hline
        & \multicolumn{2}{c|}{\textbf{w/o multiple distortion triplets}}
        & \multicolumn{2}{c|}{\textbf{w multiple distortion triplets}} \\ 
        & SRCC & PLCC & SRCC & PLCC \\ \hline
        FLIVE & 0.533 & 0.569 
              & \textbf{0.542}\pct{+1.68\%} 
              & \textbf{0.602}\pct{+5.79\%} \\ 
        SPAQ  & 0.889 & 0.892 
              & \textbf{0.893}\pct{+0.45\%} 
              & \textbf{0.897}\pct{+0.56\%} \\ 
        KonIQ & 0.853 & 0.864 
              & \textbf{0.877}\pct{+2.81\%} 
              & \textbf{0.888}\pct{+2.77\%} \\ 
        CLIVE & 0.684 & 0.730 
              & \textbf{0.725}\pct{+5.99\%} 
              & \textbf{0.767}\pct{+5.06\%} \\ \hline
    \end{tabular}}
    \label{ablation_table}
\end{table}

\section{Conclusion}
We introduced TRIQA, a novel approach to learned image quality prediction, using a training process that extracts quality related features on ranked triplets of distorted images and triplets of multiply distorted images during training. This technique enables the extraction of quality-aware features from much smaller sets of original image samples. TRIQA’s learned features demonstrate strong cross-database generalization on authentic UGC datasets, indicating that the learned representations are robust and transferable across diverse content and distortion types, and show improvement on cross-database tests. On synthetic datasets, a modified model called TRIQA-FR achieves superior performance on FR-IQA tasks compared to similarly modified SoTA NR-IQA models, without requiring additional training or fine-tuning on subjective scores. 

\vfill\pagebreak
\bibliographystyle{IEEEbib}
\bibliography{refs}

\end{document}